\documentstyle[aps,prd]{revtex}
\renewcommand{\epsilon}{\varepsilon}
\begin{document}
\preprint{}
\title{Self-interacting cosmic fluids: Particle production, causal evolution, and vacuum decay}
\author{Winfried Zimdahl\footnote{Electronic address: winfried.zimdahl@uni-konstanz.de}}
\address{Fakult\"at f\"ur Physik, Universit\"at Konstanz, PF 5560 M678
D-78457 Konstanz, Germany}
\date{\today}
\maketitle
\begin{abstract}
We discuss two different realizations of a ``deflationary'' \cite{Barrow} scenario of the early universe which imply a smooth transition from an initial de Sitter stage to a subsequent radiation-dominated Friedmann-Lema\^{\i}tre-Robertson-Walker 
(FLRW) period. 
Thermodynamically, this transition is a non-equilibrium process which may also be interpreted as the decay of an initial cosmic vacuum into relativistic matter. 
\end{abstract}

\section{Introduction}
\label{points}
This paper considers two approaches to understand the evolution of the early universe as a thermodynamical non-equilibrium process. 
The first one is closely related to the causal, second-order Israel-Stewart theory \cite{IS} for imperfect fluids, in which a quantity such as the bulk viscous pressure becomes a dynamical degree of freedom. On the other hand, an effective bulk pressure is known to be the result of cosmological particle production processes \cite{Prig,Calv,ZP1}. For ``adiabatic'' particle creation there exists a simple relation between the corresponding production rate and the equivalent effective bulk pressure. 
These properties are combined here to dynamize the rate at which the particles are produced. 
On this basis we clarify the conditions under which a ``deflationary'' evolution is consistent with causal thermodynamics.  
The second approach uses the equivalence between cosmological particle production and an effective fluid bulk pressure as well. 
But instead of dynamizing these quantities it imposes ``generalized equilibrium'' conditions on the cosmic matter. 
``Generalized equilibrium'' extends the conventional equilibrium concept of relativistic fluid dynamics to systems whithout particle number conservation \cite{ZGeq,ZiBa2}. States of generalized equilibrium may be realized in a gas with specific internal self-interactions. They are characterized by a J\"uttner distribution function, the macroscopic moments of which are not necessarily conserved, however. 
In this setting the deflationary dynamics is implied by the existence of a specific ``generalized equilibrium'' configuration, which relies on the action of suitable effective forces inside the cosmic medium.\\
The paper is organized as follows. In section \ref{deflation} we present the general relations of a deflationary model which is based on an appropriate assumption for the particle production rate. Section \ref{causal} investigates whether this model is consistent with causal, second-order thermodynamics of the Israel-Stewart type, while  
section \ref{GE} sketches its realization under the conditions of ``generalized equilibrium''.   

\section{The ``deflationary'' universe}
\label{deflation}

Both approaches to be discussed combine a conserved stress energy tensor 
\begin{equation}
T^{ik} = \rho u^{i}u^{k} 
+ \left(p + \Pi\right) h^{ik} \ 
\quad\Rightarrow\quad
\dot{\rho } + 3H \left(\rho + p + \Pi  \right) = 0 \ ,
\label{1}
\end{equation}
where $\rho $ is the energy density, $p$ is the equilibrium pressure, 
$\Pi $ is the non-equilibrium part of the scalar pressure, 
$u ^{i}$ is the fluid four-velocity, $h ^{ik} = g ^{ik} + u ^{i}u ^{k}$ is the spatial projection tensor, $H$ is the Hubble rate $H \equiv  \dot{a}/a$ and $a$ is the scale factor of the Robertson-Walker metric,   
with a non-conserved particle flow vector
\begin{equation}
N ^{i} = n u ^{i}\ ,
\mbox{\ \ \ }\mbox{\ \ \ }
N^{a}_{;a}  =  
\dot{n} + 3H n =  n \Gamma  \ , 
\label{2}
\end{equation}
where $n$ is the particle number density.  
The quantity $\Gamma = \dot{N}/N$ is the rate of change of the number 
$N \equiv  n a ^{3}$  of particles in a comoving volume $a ^{3}$. 
For $\Gamma > 0$ we have particle creation, for $\Gamma < 0$ particles are annihilated. 
From the Gibbs equation 
$T \mbox{d}s = \mbox{d} \left(\rho /n \right) 
- p \mbox{d} \left(1/n \right)$\ , 
where $T$ is the fluid temperature and $s$ is the entropy per particle, we obtain  $nT \dot{s} = - 3H\Pi - \left(\rho + p \right)\Gamma$, where we have used the balances (\ref{1}) and (\ref{2}).     
``Adiabatic'' particle production is characterized by 
$\dot{s} = 0$ \cite{Prig,Calv}: 
\begin{equation}
\dot{s} = 0 \quad\Rightarrow\quad  
\Pi = - \left(\rho + p \right)\frac{\Gamma }{3H}\ .
\label{3}
\end{equation}
The quantity $\Pi $ is determined by the rate $\Gamma $. 
Under this condition the cosmic substratum is not a dissipative fluid but a perfect fluid with varying particle number. 
Combining the field equations ($\kappa$ is Einstein's gravitational constant and $\gamma \equiv  1+p/ \rho $)  
\begin{equation}
\kappa \rho = 3 H ^{2}\ ,
\mbox{\ \ \ \ }
\dot{H} = - \frac{\kappa}{2}\left(\rho + p + \Pi  \right)
\quad\Rightarrow\quad
\kappa \Pi = - 3 \gamma H ^{2} - 2 \dot{H}
\label{4}
\end{equation} 
for a homogeneous and isotropic, spatially flat universe with condition (\ref{3}) 
yields
\begin{equation}
\frac{\Gamma }{3H} = 1 + \frac{2}{3 \gamma }
\frac{\dot{H}}{H ^{2}} 
\quad\Rightarrow\quad
\frac{H ^{\prime }}{H \left[\frac{\Gamma }{3H} - 1 \right]} 
= \frac{3}{2}\frac{\gamma }{a}
\ ,
\label{5} 
\end{equation}
where $H ^{\prime } \equiv  \mbox{d}H/ \mbox{d}a$.    
The solution of the last equation depends on an ansatz for $\Gamma /H$.  
For a dependence $\Gamma \propto \rho \propto H ^{2}$ \cite{GunzMaNe}
and for $\gamma =4/3$ we obtain 
\begin{equation}
H = 2\frac{a _{e} ^{2}}
{a ^{2} + a _{e} ^{2}}H _{e}
\quad\Rightarrow\quad
\frac{\Gamma }{3H} 
= \frac{a _{e} ^{2}}
{a ^{2} + a _{e} ^{2}}
\quad\Rightarrow\quad
N = N _{f}\left[\frac{a ^{2}}{a ^{2}+a _{e}^{2}} \right]^{3/2}\ ,
\label{6}
\end{equation}
where we have choosen the constants such that 
$\dot{H}_{e} = - H _{e}^{2}$, i.e, $\ddot{a}>0$ for $a<a _{e}$ and 
$\ddot{a}<0$ for $a>a _{e}$. 
$N _{f}$ may be identified with the ``final'' particle number in the presently observed universe.  
$H$ starts with a constant value $H _{0}=2H _{e}$ at $a \ll a _{e}$ and then ``deflates''  towards the typical $H \propto a ^{-2}$ behaviour of a radiation dominated universe for $a \gg a _{e}$.  
This Hubble rate has also been obtained in the context of phenomenological approaches to cosmological vacuum decay \cite{GunzMaNe,LiMa}. 
The corresponding connection is easily established if we perform the split 
$\rho = \rho _{1}+\rho _{2}$ with 
\begin{equation}
\rho _{1} = \frac{3 H _{e}^{2}}{2 \pi }m _{P}^{2}
\left(\frac{a}{a _{e}} \right)^{2}
\left[\frac{a _{e}^{2}}{a ^{2} + a _{e}^{2}} \right]^{3}\ , 
\mbox{\ }\mbox{\ }\mbox{\ }
\rho _{2} = \frac{3 H _{e}^{2}}{2 \pi }m _{P}^{2}
\left[\frac{a _{e}^{2}}{a ^{2} + a _{e}^{2}} \right]^{3}\ ,
\label{7}
\end{equation}
where $m _{P}$ is the Planck mass ($\kappa = 8 \pi /m _{P}^{2}$). 
The part $\rho _{2}$ corresponds to a cosmological term which decays from 
a finite, nonzero initial value at $a=0$ as $a ^{-6}$ for 
$a \gg a _{e}$, while 
the part $\rho _{1}$ describes relativistic matter with 
$\rho _{1}\rightarrow 0$ for $a \rightarrow 0$. 
According to (\ref{6}) the initial particle number $N _{0}$ is zero, i.e., all the (ultrarelativistic) particles are produced  at the expense of the decaying vacuum energy density $\rho _{2}$. 
In the following we discuss two situations  which may give rise to 
such a scenario.

\section{Causal evolution}
\label{causal}
The first case is based on the second-order expression for the entropy-flow vector \cite{IS}
\begin{equation}
S^{a} = sN^{a} 
- \frac{\tau\Pi^2}{2\zeta T} u^{a} \ , 
\label{8}
\end{equation}
where 
$\tau$ is a  relaxation time and   
$\zeta$ an effective coefficient of  bulk
viscosity. 
The requirement $S^{a}_{;a}\geq 0$ implies a relaxation equation for $\Pi $. 
Via (\ref{3}) this is equivalent to a corresponding equation for $\Gamma $, i.e.,  
the latter quantity is a dynamical degree of freedom in our approch. 
It is important to realize that the relaxation time $\tau $ here characterizes the creation process but {\it not} the deviation from local equilibrium as in the standard second-order theory where the particle number is preserved \cite{Zpreprint}. 
The idea is now to check under which condition such an equation for $\Gamma $ 
is consistent with the dynamics (\ref{6}). 
It turns out that the positivity requirement for the relaxation time restricts the admissible dynamics  to  \cite{Zpreprint}
\begin{equation}
H < \frac{4}{7}\left[1 + \sqrt{1 + \frac{21}{2}c _{b}^{2}} \right]H _{e}\ , 
\mbox{\ \ \ }\mbox{\ \ \ }
c _{b}^{2} = \frac{4}{3}\frac{\zeta }{\rho \tau } \leq \frac{2}{3}\ ,
\label{9}
\end{equation}
where the last inequality follows from the condition that the propagation velocity $c _{s}^{2}   + c _{b}^{2}$ \cite{Roy}  (where $c _{s}$ is the adiabatic sound velocity with $c _{s}^{2}   = 1/3$ in the present case) be smaller than the velocity of light. 
As long as the non-equilibrium part $c _{b}^{2}$ lies in the range $2/3 \geq c _{b}^{2} >1/2$, the whole scenario (\ref{6})   starting with $H _{0}=2H _{e}$ at $a=0$ is    
compatible with general causality requirements for effective bulk viscous fluids. 
For $c _{b}^{2}<\frac{1}{2}$, on the other hand, the initial de Sitter stage is not a solution of the causal evolution equation. 
Even for $c _{b}^{2}  \ll 1$, however, where condition (\ref{9}) becomes 
$H<\frac{8}{7}H _{e}$, the applicability of (\ref{6}) still starts within the phase of accelerated expansion ($H>H _{e}$). 
We emphasize that this restriction represents an inherent self-limitation of the theory concerning the magnitude of the effective bulk pressure $\Pi $. 
By contrast, the standard second-order theory for dissipative fluids with conserved particle numbers does {\it not} provide limits for the thermodynamic fluxes. 
In the present context this limitation precises the conditions under which an effectively imperfect fluid picture of the inflationary universe is consistent with causal, second-order thermodynamics of the Israel-Stewart type.

\section{``Generalized'' equilibrium}
\label{GE}
In the second case the deflationary model of section \ref{deflation} is a consequence of the generalized equilibrium condition for a self-interacting gas universe. 
This condition is 
\begin{equation}
\pounds _{_{\frac{u _{a}}{T}}} g _{ik}   
\equiv  \left(\frac{u _{i}}{T} \right)_{;k} 
+ \left(\frac{u _{k}}{T} \right)_{;i}
= \frac{{\rm 2} H}{T}
\left[g _{ik} + \frac{a _{e}^{{\rm 2}}}{a ^{{\rm 2}}
+a _{e}^{{\rm 2}}} u _{i}u _{k}\right] \ , 
\label{10}
\end{equation}
with $H$ form (\ref{6}) . 
It changes continously from the initial expression \cite{ZiBa2}
\begin{equation}
\pounds _{_{\frac{u _{a}}{T}}} g _{ik}  
= \frac{{\rm 2} H}{T}
h _{ik} 
\mbox{\ \ } 
{\rm for} 
\mbox{\ \ } 
a \ll a _{e}
\mbox{\ \ }
{\rm to}
\mbox{\ \ }
\pounds _{_{\frac{u _{a}}{T}}} g _{ik}  
= \frac{{\rm 2} H}{T}
g _{ik}
\mbox{\ \ } 
{\rm for} 
\mbox{\ \ } 
a \gg a _{e}\ . 
\label{11}
\end{equation}  
The last (conformal Killing vector) condition is known to follow from Boltzmann's equation as ``global'' equilibrium condition for the one-particle distribution function 
$f \propto \exp{\left[-E/T \right]}$ 
of a classical ultrarelativistic gas. Here $E$ is the particle energy 
$E \equiv  -u _{i}p ^{i}$ 
where $p ^{i}$ is the particle four-momentum.   
Generalizing the methods of \cite{ZiBa2} one can show that condition (\ref{10})  follows in a similar way as equilibrium condition from the Boltzmann equation if the gas particles are moving in an additional effective 
force field $F ^{i}$, i.e., 
$\mbox{D}p ^{i}/\mbox{d}\lambda = F ^{i}$ (where $\lambda $ is a parameter along the particle worldline) instead of geodesic particle motion 
($F ^{i}=0$) as it is assumed in the derivation of the second condition in (\ref{11}). 
In the case of interest here the force which is necessary to maintain an equilibrium distribution is given by 
$
u _{i}F^{i}= - H E ^{2}
a _{e}^{2}\left[a _{e}^{2}+a ^{2} \right]^{-1}$. 
This force is self-consistently exerted by the cosmic medium on each of its microscopic constituents, i.e., it represents a self-interaction of the cosmic substratum. 
If such a self-interaction exists, it makes the individual particle energies 
decay at a rate $\dot{E}/E$ which exactly coincides with the cooling rate 
\begin{equation}
\frac{\dot{T}}{T} \equiv  \frac{1}{2}T
u ^{i}u ^{k} \pounds _{_{\frac{u _{a}}{T}}} g _{ik}  
\quad\Rightarrow\quad
T \propto 
\left[\frac{a _{e}^{{\rm 2}}}
{a ^{{\rm 2}} + a _{e}^{{\rm 2}}} \right]^{{\rm 1}/{\rm 2}} \ ,
\mbox{\ \ \ } 
\frac{E}{T} = {\rm const}\ .
\label{12}
\end{equation} 
The quantities $E$ and $T$  start with finite constant initial values at $a=0$ and approach the familiar dependence 
$E \propto T \propto a ^{-1}$ for $a \gg a _{e}$.  
The relations (\ref{6}), (\ref{10})  and (\ref{12}) establish an exactly solvable model, both macoscopically and microscopically, of a deflationary gas universe in ``generalized equilibrium''.

\acknowledgments
This paper was supported by the Deutsche Forschungsgemeinschaft.


\end{document}